\documentclass{JINST}
\let\ifpdf\relax
\usepackage{graphicx}
\usepackage{amsmath}   
\usepackage{subfigure}
\usepackage{wasysym}
\usepackage{lineno}
\usepackage{cite}

\title{Design and operation of ARGONTUBE: a 5 m long drift liquid argon TPC}

\author{A. Ereditato$^a$, C.C. Hsu$^a$, S. Janos$^a$, I. Kreslo$^a$, M. Messina$^a$\footnote{Currently at Columbia University, Department of Physics, New York, NY, USA.}, C. Rudolf von Rohr$^a$, B. Rossi$^a$\footnote{Currently at Princeton University, Department of Physics, Princeton, NJ, USA.}, T. Strauss$^a$,  M.S. Weber$^a$, M. Zeller $^a$\thanks{Corresponding author.} \\
\llap{$^a$}Laboratory for High Energy Physics, Albert Einstein Center for Fundamental Physics (AEC), University of Bern,\\
Sidlerstrasse 5,3012 Bern, Switzerland.\\
  E-mail: \email{marcel.zeller@lhep.unibe.ch}}

\abstract{
The Liquid Argon Time Projection Chamber (LArTPC) is a prime type of detector for future large-mass neutrino observatories and proton decay searches.
In this paper we present the design and operation, as well as experimental results from ARGONTUBE, a LArTPC being operated at the AEC-LHEP, University of Bern.
The main goal of this detector is to prove the feasibility of charge drift over very long distances in liquid argon.
Many other aspects of the LArTPC technology are also investigated, such as a voltage multiplier to generate high voltage in liquid argon (Greinacher circuit), a cryogenic purification system and the application of multi-photon ionization of liquid argon by a UV laser.
For the first time, tracks induced by cosmic muons and UV laser beam pulses have been observed and studied at drift distances of up to 5 m, the longest reached to date.
}


\begin{document}

\section{Introduction}

Neutrino oscillations have been established with compelling evidence in experiments conducted in the last two decades with solar, atmospheric, accelerator and reactor neutrinos leading to the determination of the neutrino mass eigenvalue differences and of the three mixing angles \cite{PDG} of the PMNS mixing matrix  \cite{pontecorvo, MNS}.


The next generation of neutrino oscillation experiments will aim at providing precision measurements of $\theta_{13}$, investigating possible CP violating effects in the lepton sector and determining the hierarchy of the mass eigenvalues.
To achieve the necessary sensitivity, large samples of events are needed, which can only be obtained with a synergic use of high flux neutrino beams and very large mass detectors.
A prime technology for such detectors are Liquid Argon Time Projection Chambers (LArTPC) \cite{Rubbia, ICARUSproposal}, also well suited for sensitive searches for proton decay.
A design option with a single cylindrical volume of 70~m diameter and 20~m in height, GLACIER, is discussed in \cite{GLACIER1, GLACIER2, GLACIER3} and references therein.

TPCs are three-dimensional tracking devices with excellent calorimetric and particle identification capabilities \cite{zeller, rossi, larn}.
In the TPC, charged particles produce ionization electrons, which are drifted towards readout planes by an electric field.
The latter is obtained by a cathode and field shaper rings installed at fixed pitch and biased with increasing high voltage.
The reconstruction of particle tracks is realized by recording the electric signals generated by the drifted electrons on the readout planes, defining the transverse coordinates.
The longitudinal space coordinate is obtained from the measurement of the drift time.
Given its high density of 1.4~g/cm$^{3}$, liquid argon is well suited for neutrino physics TPCs, acting simultaneously as target and active medium.

At the AEC-LHEP we conduct a vigorous R\&D program \cite{zeller, rossi, badhrees} to further develop the LArTPC technique for use in future neutrino and matter instability search experiments.
One of the main challenges addressed is the achievement of very long drift distances.
Before our measurements presented here, the longest drift length reached was 1.5~m, obtained by the ICARUS collaboration\cite{ICARUS}.
Long drift distances require the liquid argon to be very pure, in order to minimize the charge depletion due to the attachment of electrons to impurities having negative electron affinity.
Typical impurities are O$_{2}$, H$_{2}$O and CO$_{2}$.
Liquefied noble gases themselves have positive electron affinity.
Furthermore, the electric field needs to be in the order kV/cm and, therefore, the voltage potential to be applied  over several meters is very high (up to MV).
At lower electric field values, the usable charge at the ionization point becomes too small due to the recombination of the electron-ion pairs\cite{Imel}.
Low drift field values also decrease the electron drift speed, thus increasing the drift time and resulting in higher electron attachment.
Diffusion may also significantly affect the position resolution of the TPC \cite{diffusion, diffusion2, diffusion3, atrazhev} at long drift times.

In this paper we report on the design and operation of ARGONTUBE, a liquid argon TPC with which a drift distance as long as 5~m was achieved as discussed in the following.
The design of ARGONTUBE is presented in Section~\ref{argontube}.
Important technological advances include a cryogenic purification system presented in Section~\ref{cryo}, as well as a voltage multiplier to generate high voltage in situ in the TPC (Greinacher circuit) described in Section~\ref{greini}. 
To study the detector performance, cosmic muons and controlled ionization tracks induced by a UV laser were used, a novel technology proposed and mastered by our group.
The laser system is described in Section~\ref{laser}.
The operation of the detector and the results from the performance measurements are reported in Section~\ref{results}.

\section{The ARGONTUBE detector}
\label{argontube}
\begin{figure}
\centering	
\includegraphics[width=1\linewidth]{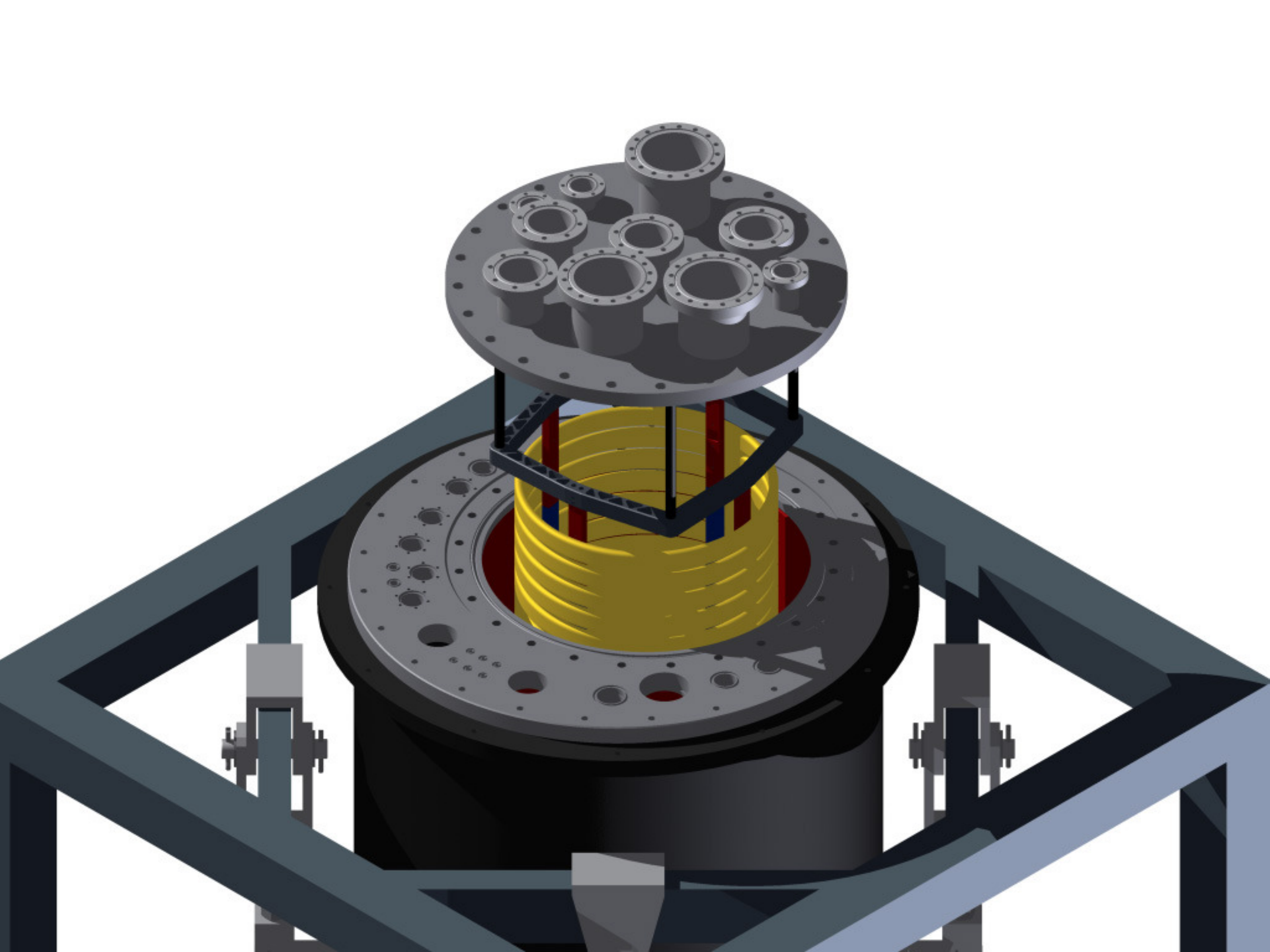}
\caption{Drawing of the ARGONTUBE cryostat with the inner vessel and the top flange with all the feedthroughs. A part of the field shaping rings and some details of the fastening system to the top flange are shown. Only a few rings out of 125 are visible in this picture.}
\label{fig:TPC}
\end{figure}

ARGONTUBE is a LArTPC of 5~m length in drift direction and 40~cm in diameter.
The cryogenic system consists of an outer vacuum insulated cryostat, into which an inner vessel (5.6 m long and with 50 cm diameter) is inserted.
The outer vessel is vacuum insulated, with 50 layers of super insulation added to minimize the heat input from thermal radiation.
For operation, both volumes are filled with liquid argon.
The outer volume acts as a bath, keeping the inner volume cold and reducing boiling.
The heat input into the outer volume is compensated by the boiling of the argon.
In Figure \ref{fig:TPC} a drawing of the ARGONTUBE detector is shown, with the dewar, the inner vessel and a part of the field cage.
All connections to the inner vessel are done with CF-flanges and the top flange is sealed with indium.

\subsection{Electric field simulation and field cage design}
\label{esim}

\begin{figure}
\centering
\mbox{\subfigure{\includegraphics[width=0.55\linewidth]{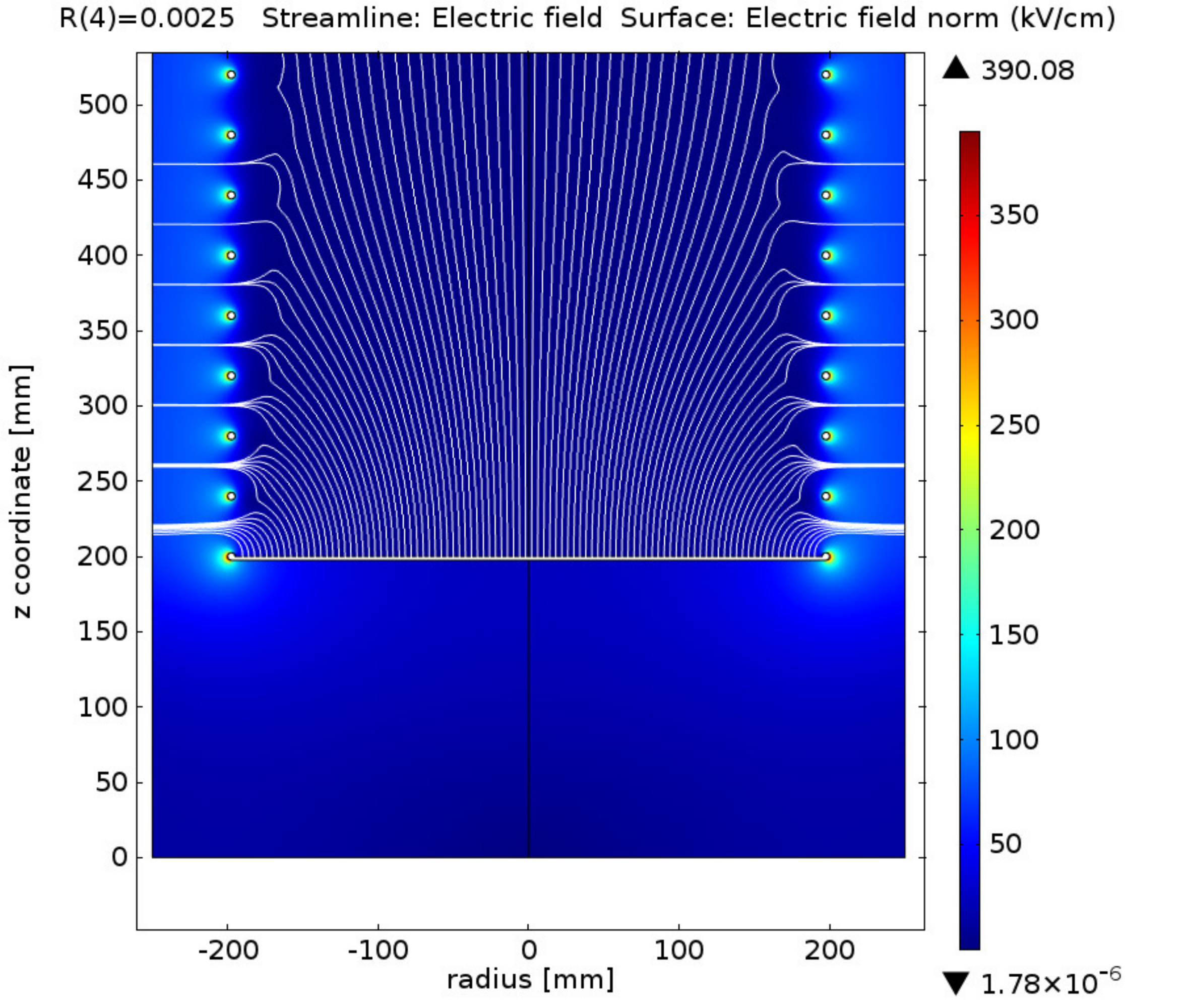}
\quad
\subfigure{\includegraphics[width=0.55\linewidth]{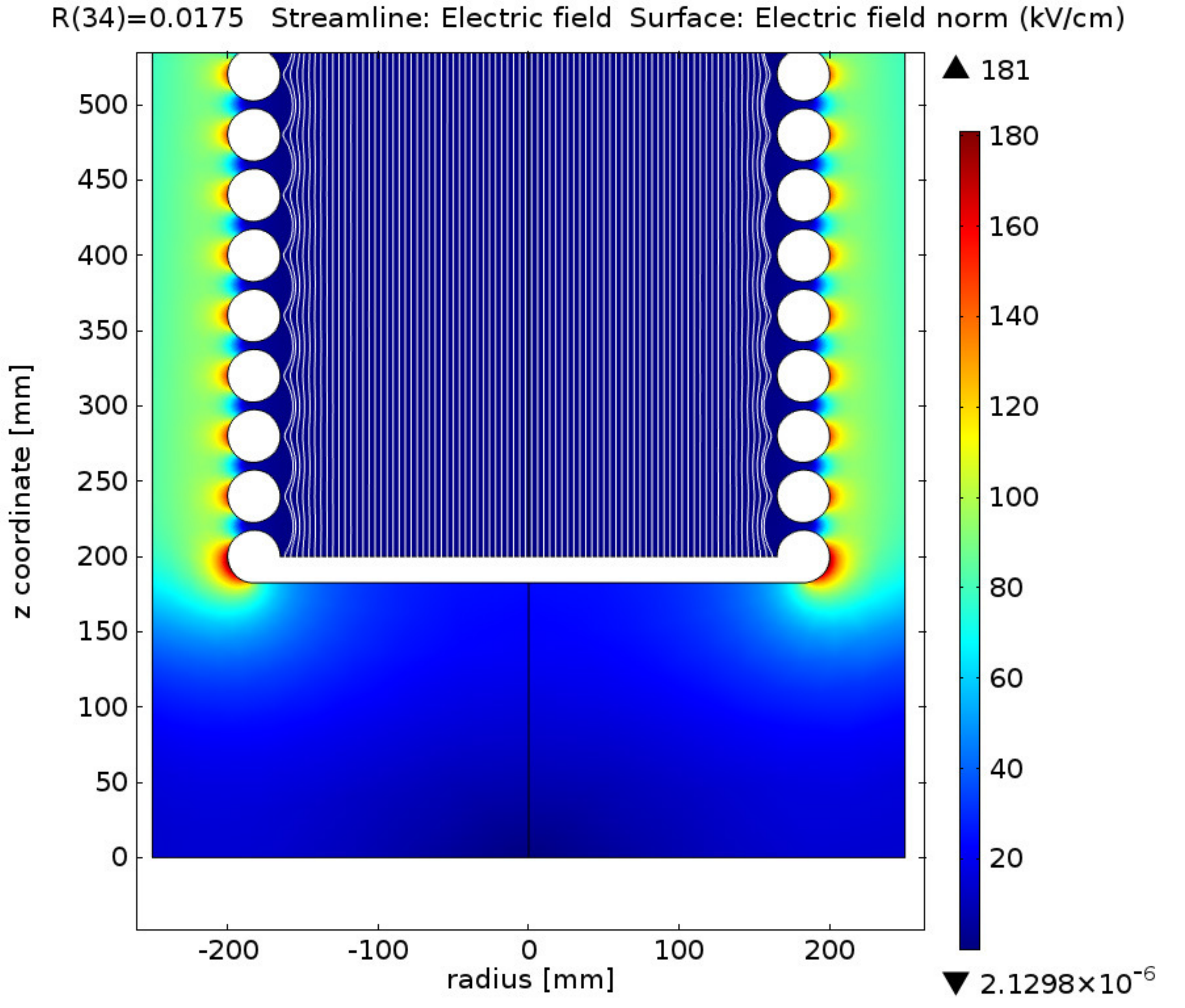} }}}
\caption{Different field shaping ring diameters were simulated. On the left picture the field-shaping ring diameter is 5~mm and on the right picture 35~mm. The electric field strength is given by the color range and the electric field lines in the drift volume are shown in white. Note that the color scales are different for the two pictures.}
\label{fig:E-Pot}
\end{figure}

The electric field has a design value of 1 kV/cm, which results in a potential of 500~kV at the cathode at 5~m distance from the readout planes.
A detailed FEM (Finite-Element-Method) simulation of the electric field was made for ARGON\-TUBE in order to design a drift-field in the TPC as uniform as possible.
Furthermore, the electric field between the field shaping rings and the cryostat wall was minimized.
In the design of the field shaping rings also the mechanical stability was optimized while minimizing the weight.
We used the COMSOL\footnote[1]{web page: http://www.comsol.com/} code to simulate the electrostatic field inside the TPC.
A 2D-axial-symmetric simulation was made since our detector is rotation-symmetric in a first approximation.
Different cross section forms and sizes of field-shaping rings were studied.
In a first simulation, round field-shaping rings with different diameters were simulated.
In Figure~\ref{fig:E-Pot} the electric field strength and the electric field lines are shown for 5~mm and 35~~mm ring cross section diameters.
Only the lowest part of the TPC is shown, where the strongest field to the wall and the strongest disturbance to the drift field occur.
For 5~mm rings the highest field is 0.39~MV/cm and the electric field lines are rather non-uniform; for 35~mm rings the highest field is reduced to 0.18~MV/cm and the electric field lines are much more uniform.

\begin{figure}
\centering	
\includegraphics[width=0.8\linewidth]{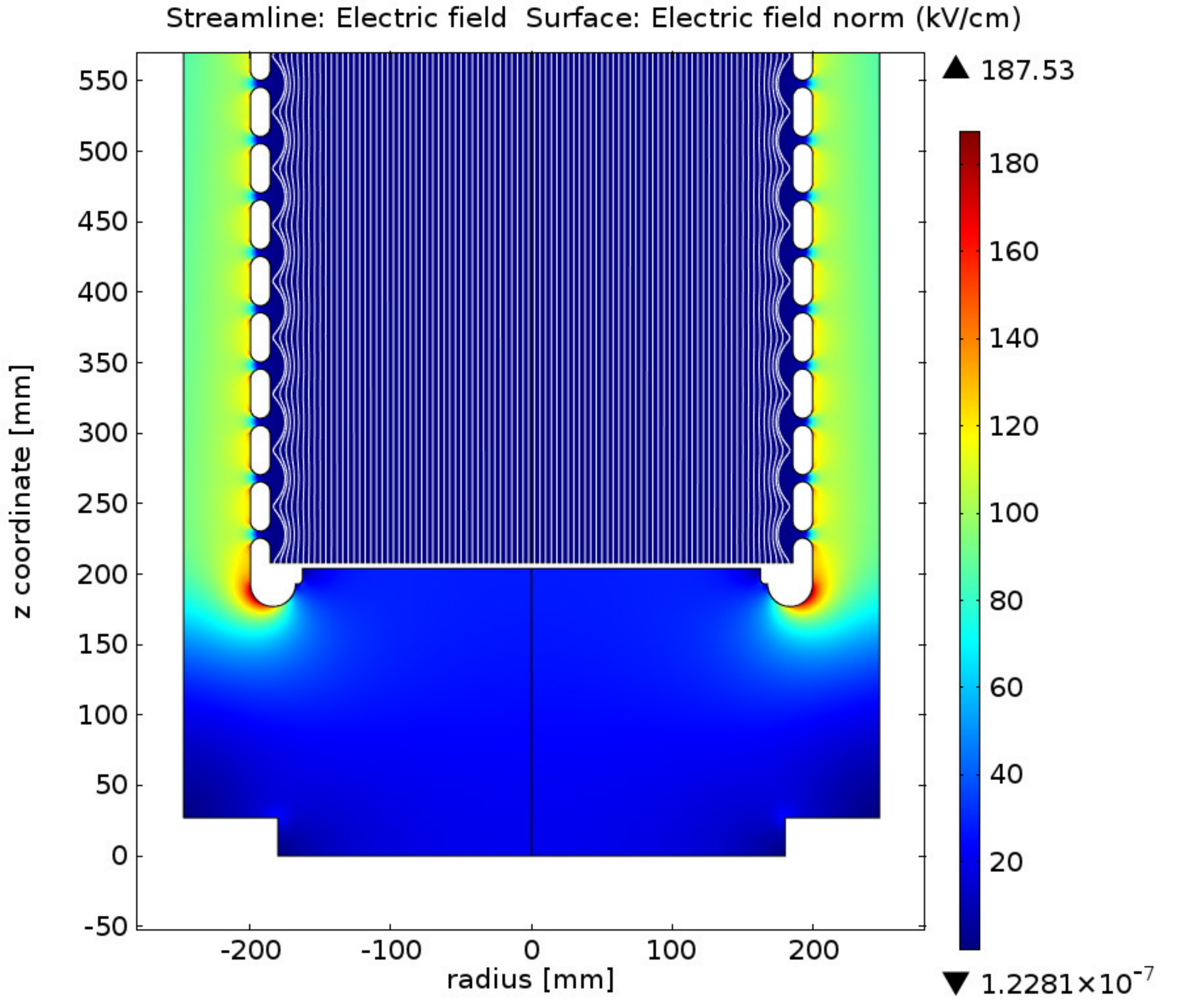}
\caption{Final design of the TPC field cage. The electric field strength is given by the color range and the electric field lines in the drift volume are shown in white. }
\label{fig:E-lines}
\end{figure}

In order to minimize the weight of the field-shaping rings and to increase the drift volume a race-track cross-section shape was then simulated, as can be seen in Figure \ref{fig:E-lines}.
We find that the optimal race-track shape has a thickness of 1.6~cm, a height of 3.4~cm and an inner radius of 8~mm.
In this configuration electric field values up to 0.187~MV/cm are obtained between the cathode plate and the side wall, which is slightly higher compared to the round rings.
On the other hand, the active volume is increased and the weight of the field cage is significantly reduced.
The maximal electric field is still much lower than the breakdown strength of liquid argon, expected to be 1.1-1.4~MV/cm \cite{swan}. 
The optimal pitch for placing the field-shaping rings is 4~cm, with a space between the rings of 5~mm.
This results in 125 rings mounted in a column to build the field cage, each of them having an outer diameter of 40~cm.
With the cathode plate at a design voltage of 500~kV the potential difference between two rings is 4~kV, providing the design drift field of 1~kV/cm.
In Figure \ref{fig:E-lines} also the resulting electric field strength and the electric field lines in the TPC are shown.
The field variations in the middle of TPC are smaller than 0.1\% and at 150~mm distance from the center the field variations are less than 0.5\%, which is well within the surface readout of 20$\times$20~cm$^2$. 

The Greinacher high voltage circuit presented in Section~\ref{greini} is placed inside the field cage in order to minimize the electric field at the surface of its small structures with radius less than one millimeter.
Inside the TPC the drift-field is only slightly disturbed by the Greinacher circuit. 

\begin{figure}
\centering
\mbox{\subfigure{\includegraphics[width=0.5\linewidth]{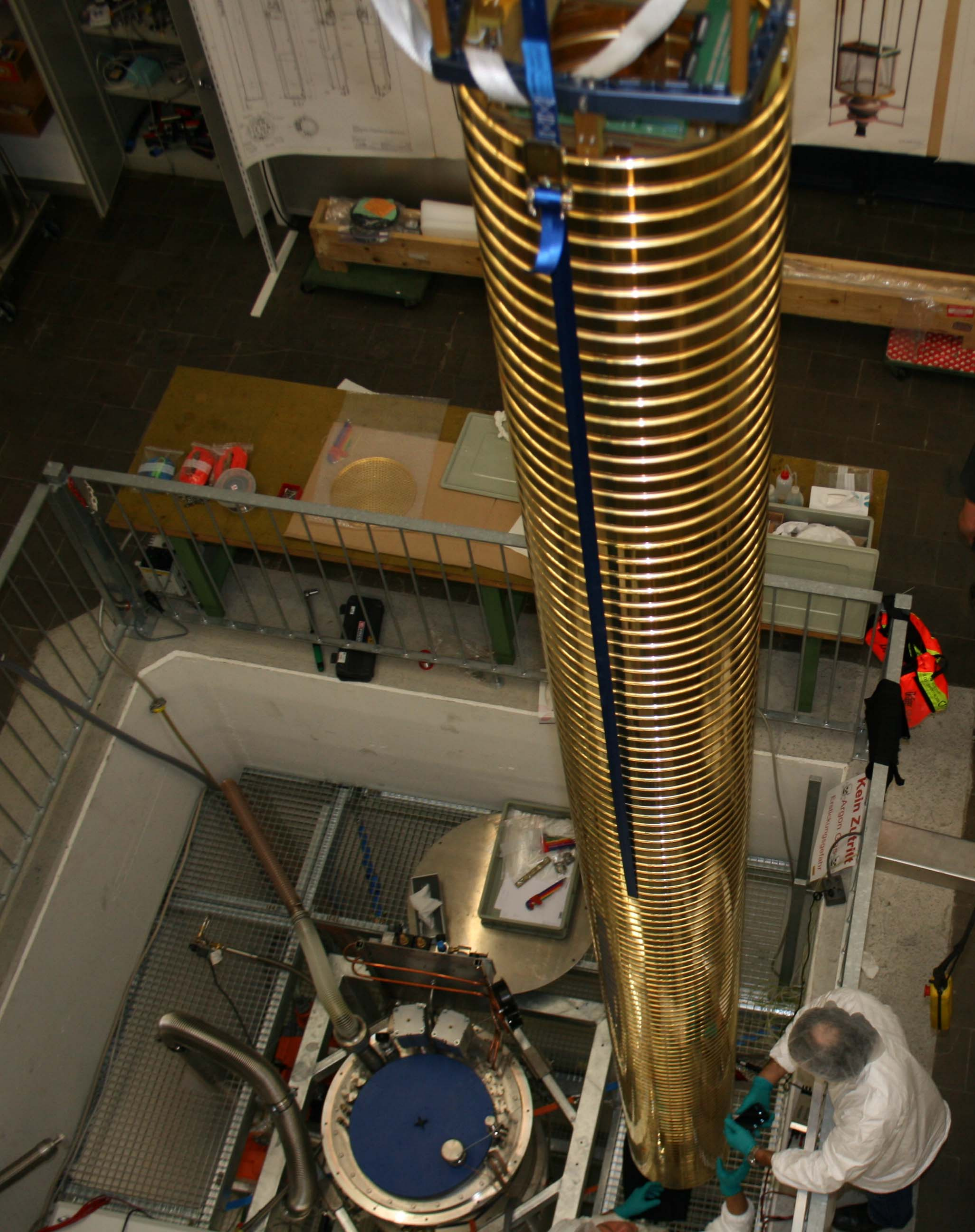}
\quad
\subfigure{\includegraphics[width=0.425\linewidth]{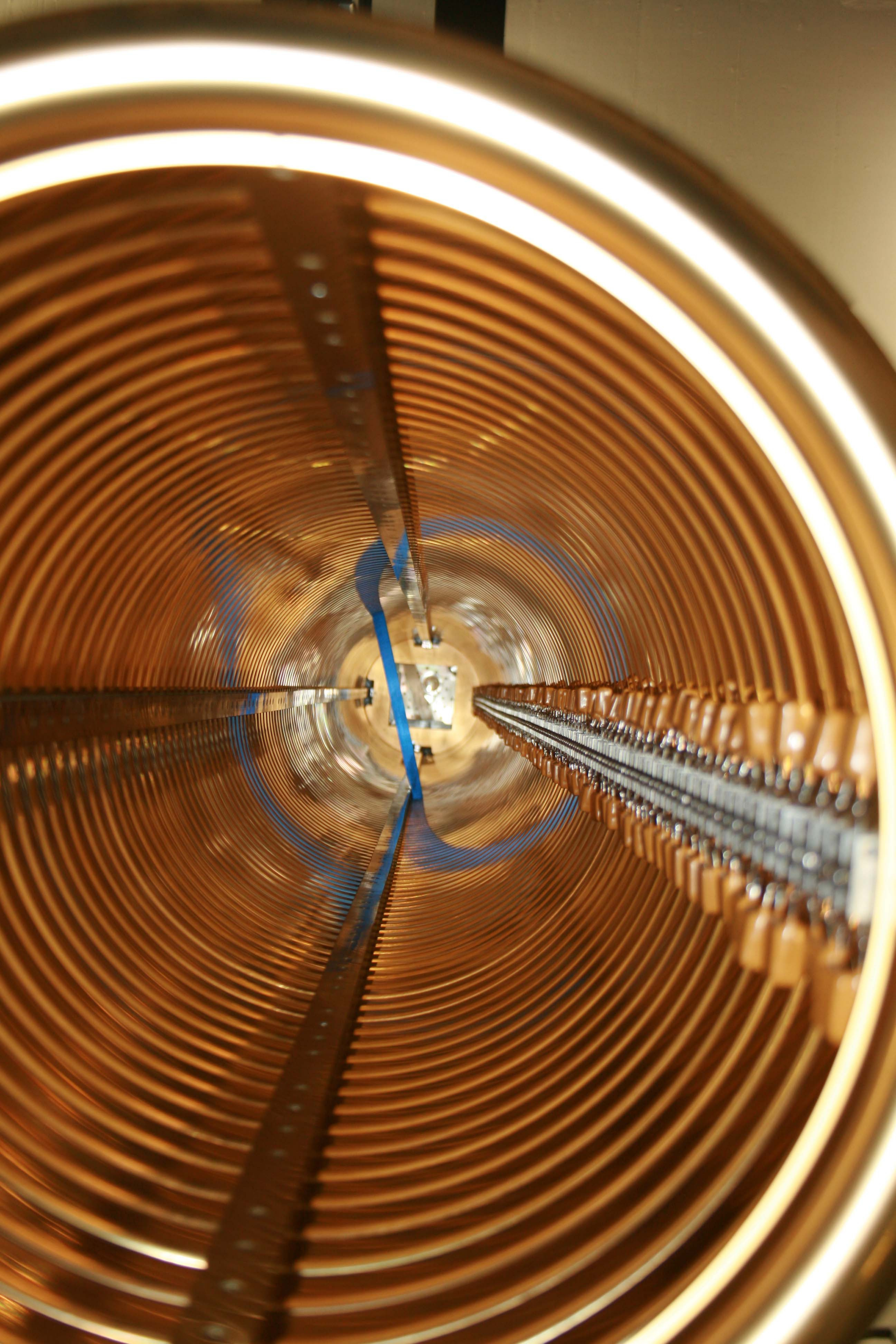} }}}
\caption{Left: Picture of the TPC with the 125 field shaping rings. Right: View inside the TPC with the mounted Greinacher/Cockroft-Walton circuit visible on the side. }
\label{fig2}
\end{figure}

In Figure \ref{fig2} the assembled TPC is shown with 125 field-shaping rings.

These are made of solid aluminum, polished and gold plated to obtain a very clean and inert surface.
The total weight of the field cage with the wire planes is about 250~kg.
The supporting structure holding the field cage is made out of PAI (Polyamidimid + Ti0$_2$ + PTFE), which keeps its mechanical strength from -200~$^\circ$C to +250~$^\circ$C, allowing for operation in liquid argon.
Other characteristics of PAI are its extremely low thermal expansion for a plastic (30$\times$10$^{-6}$~K$^{-1}$ at room temperature), very high electric resistivity (2$\times$10$^{15}$~$\Omega$m), very high mechanical stability (tensile strength of 218~MPa  at -196$^\circ$C), high stiffness, excellent impact resistance and very good radiation hardness.
The whole structure is hanging on four pillars of PAI fixed to the top flange, as can be seen in Figure \ref{fig:TPC}.
By lifting up the top flange the whole field cage can be extracted for maintenance and repair. 
To obtain a very smooth surface for best cleanliness and high voltage performance also the inner surface of the inner vessel was electro-polished.

\subsection{Cryogenic purification system}
\label{cryo}
\begin{figure}
\centering
\mbox{\subfigure{\includegraphics[width=0.4\linewidth]{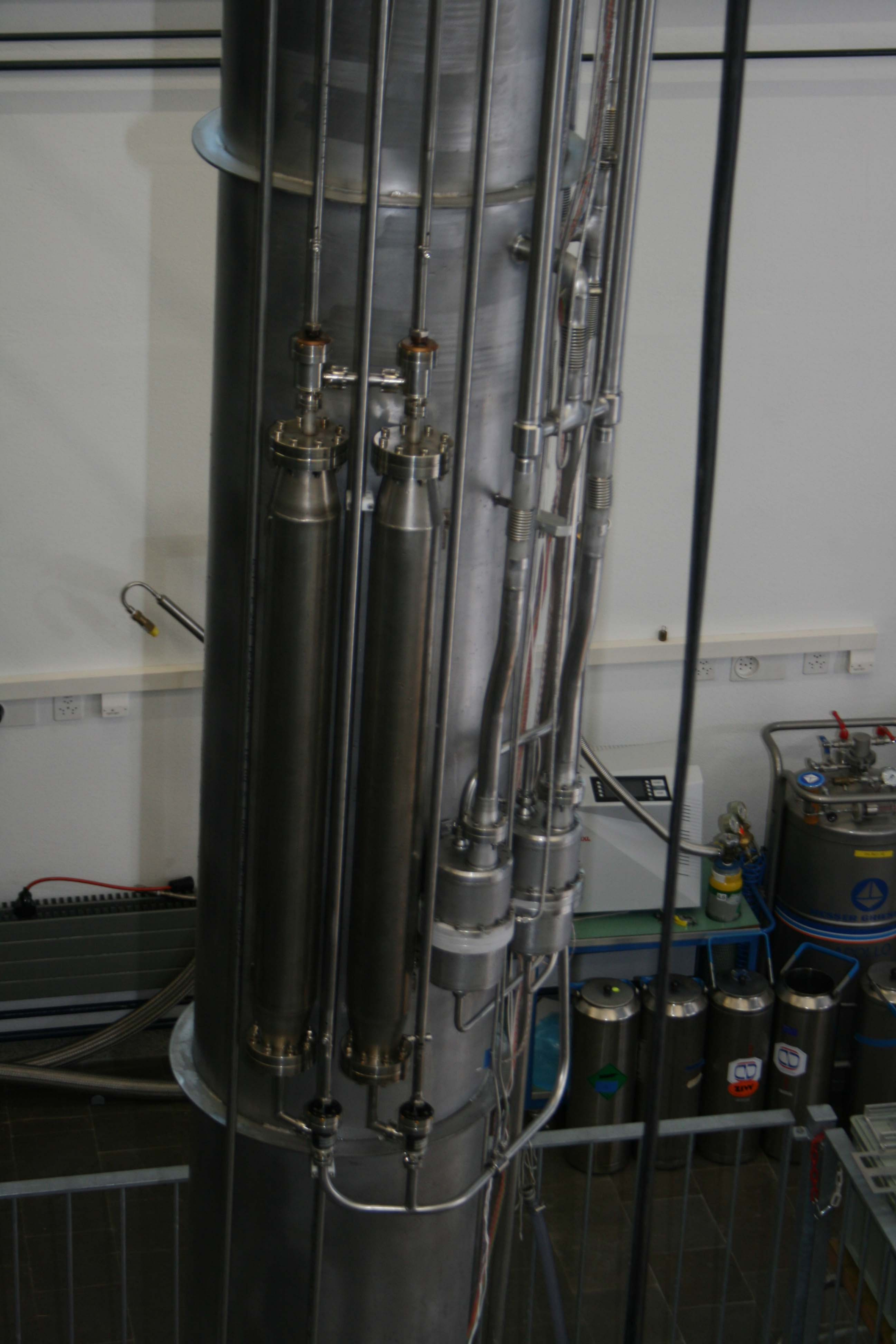}
\quad
\subfigure{\includegraphics[width=0.4\linewidth]{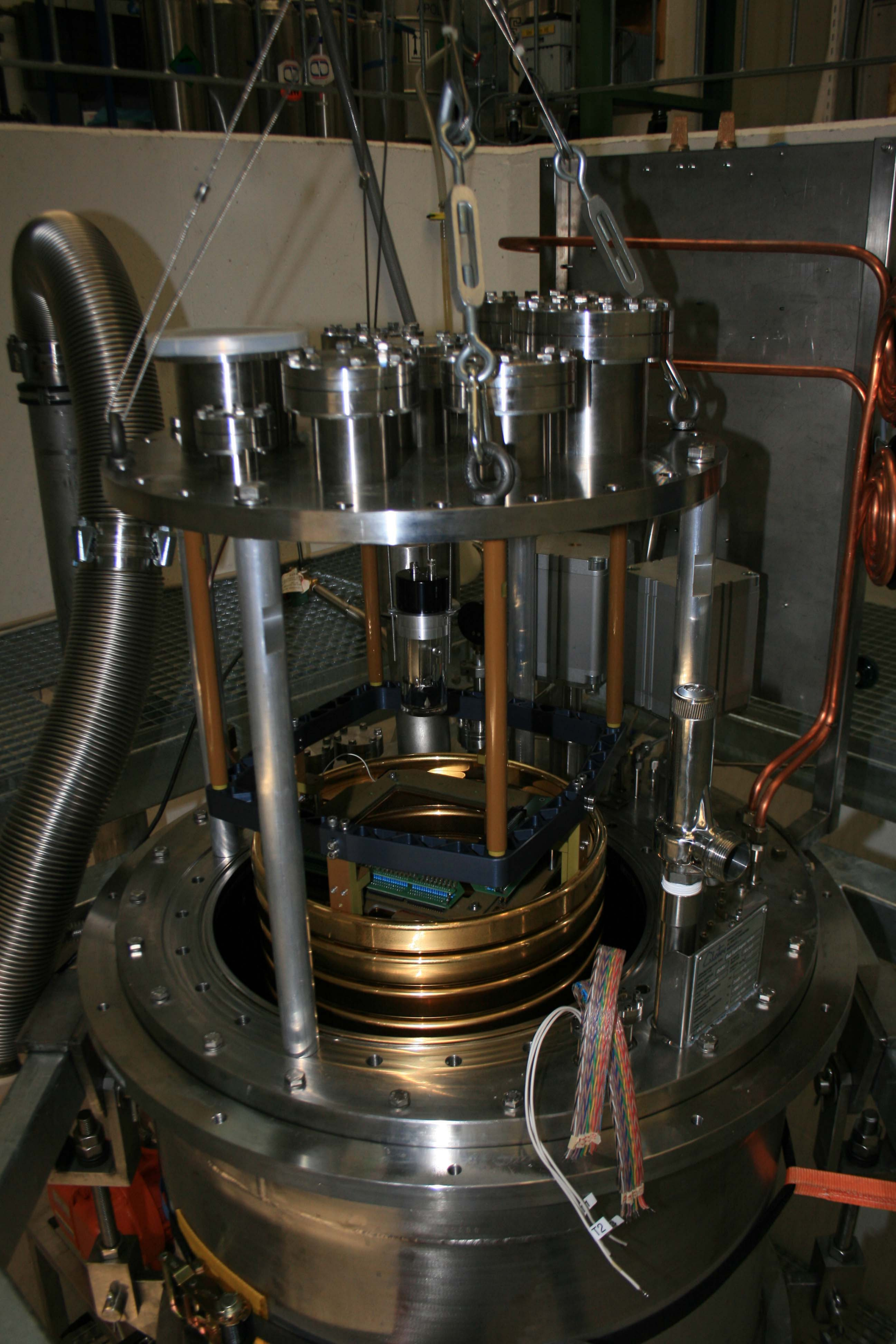} }}}
\caption{Left: Picture of the inner vessel where the Oxygen-trapping filters and the bellow pump are shown. Right: Picture of the top flange with all feed troughs, field shaping rings, some details of the fastening system to the top flange and two photomultipliers used as trigger.}   
\label{fig1}
\end{figure}

The commercial liquid argon used for ARGONTUBE is certified to have oxygen-like impurities of at most 20~ppm and is stored in a 5000~liter storage tank connected to the detector through vacuum isolated transport lines with super isolation.
Since the required purity is of the order of 0.1~ppb, additional purification is required.
As a first measure, the liquid argon is filled through cryogenic oxygen trapping filters manufactured by  CRIOTEC \footnote[2]{CRIOTEC Impianti S.r.l, Via Francesco Parigi 4 - zona industriale CHIND, 10034 CHIVASSO (TO) - ITALY.} based on activated copper powder.
In a chemical reaction, oxygen and water are bound to the copper.
To further increase the purity of the liquid argon and to compensate possible outgassing of detector components in the warm phase, the inner volume is equipped with two cryogenic pumps shown in Figure \ref{fig1} that circulate liquid argon through two additional filters. The filters used for the recirculation are of the same type as used for the filling.
We have chosen a bellow pump system which provides a recirculation speed of 300~l/min of liquid
with minimal electromagnetic interference to the TPC readout system and very limited heat input.
The bellow pumps are driven by  pressurized nitrogen gas.

\subsection{High voltage system}
\label{greini}
\begin{figure}
\centering	
\includegraphics[width=1\linewidth]{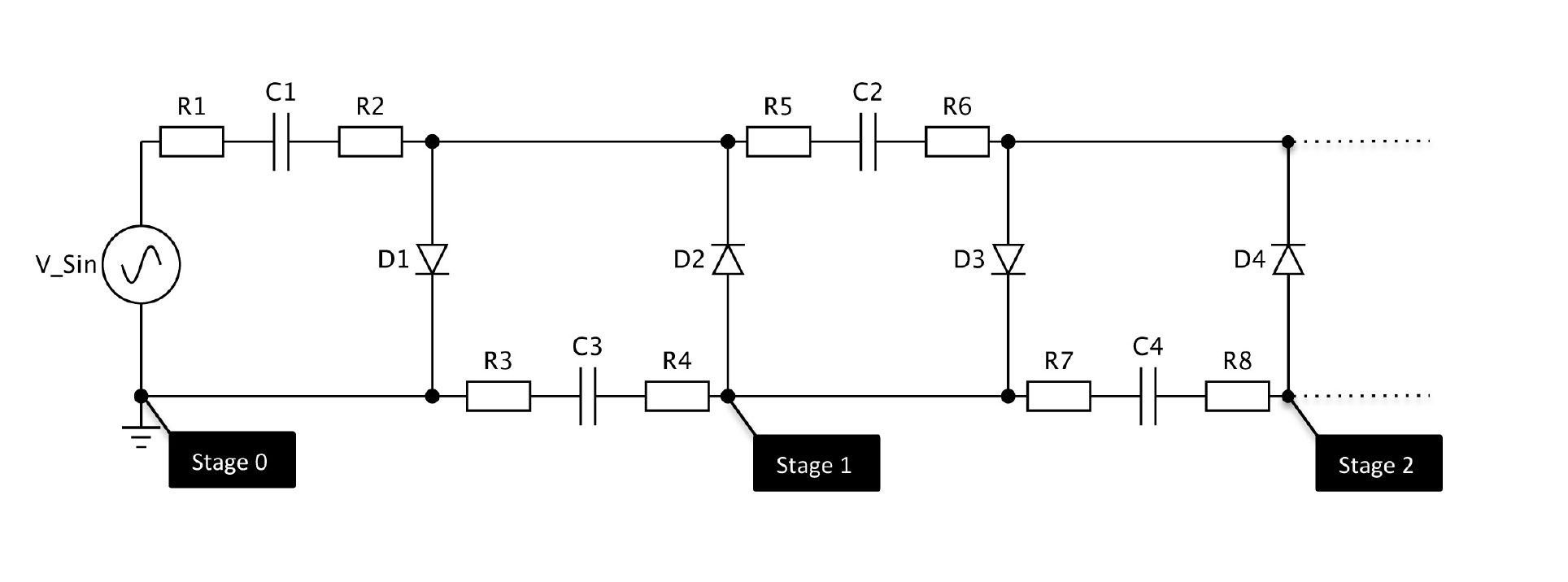}
\caption{Wiring diagram of the Greinacher circuit for negative high voltage. Two stages of this circuit are shown, in ARGONTUBE 125 stages are mounted.}
\label{fig:wdgs}
\end{figure}

In a long drift TPC, as in the case of ARGONTUBE, the high voltage needed to generate the drift field is critical.
Given the difficulty to feed the required voltage (500~kV) into the liquid argon without provoking discharges, we developed a voltage multiplier based on a Greinacher (also know as Cockcroft-Walton) circuit consisting of a chain of voltage multiplying stages.
An AC charging voltage is applied to the circuit in which the voltage is rectified and multiplied by a factor corresponding to the number of stages.
In Figure \ref{fig:wdgs} a diagram of the Greinacher circuit with two stages is depicted.
The Greinacher circuit in ARGONTUBE is installed in-situ in the liquid argon.
We have chosen a configuration in which each stage of the Greinacher circuit is biasing one field-shaping ring.
With a distance between each ring of 4~cm and an input AC voltage of 2~kV (4~kV peak-to-peak) the design field of 1~kV/cm is reached. 
In total 125 stages are installed resulting in an overall voltage of 500~kV at the cathode. 
The actual implementation of the circuit can be seen in Figure \ref{fig2}, where the capacitors and the diodes are also visible.
The capacitors\footnote[3]{Part Nr. 7565K803K402LEX-A, NOVACAP, 25111 Anza Drive, Valencia, CA 91355, USA.} hold a voltage of 4~kV and in liquid argon at 87~K have a capacitance of 158~nF. The diodes\footnote[4]{Part Nr. M160UFG, VMI Voltage Multipliers Inc., 8711 W. Roosevelt Ave., Visalia, CA 93291, USA.} hold a reverse voltage of 16~kV and open at 15~V.
To limit the current in the case of an accidental discharges each capacitor is connected via two 15~kOhm resistors.
This slightly increases the charging time, which however is not a significant issue to the operation.
The input AC voltage is provided by a regulated power supply with 0-4~kV peak-to-peak at a frequency of 50~Hz.
It consists of a signal generator that makes a 0-4~V sinusoidal signal amplified with a transformer by a factor of 1000.
The input voltage and the input current are constantly measured to monitor the charging process.
The circuit takes about 30 minutes to charge, while it discharges by less than 1$\%$ in one hour.

\subsection{Charge readout,  DAQ and trigger}
\label{daq}
Two readout planes are mounted on top of the TPC each of them consisting of 64 wires.
The electrons drifting in the TPC and approaching the wire planes induce a signal on a first plane; they are then collected by a second plane.
The wires are made of a Be-Cu alloy, have a diameter of 125~$\mu$m and are 20~cm long.
Wires are mounted with a pitch of 3~mm yielding a readout surface of 20~cm by 20~cm.
Both wire planes are equipped with decoupling capacitors in order to bias their DC potential.
The charge signal is fed to pre-amplifiers \cite{rossi} placed on the top flange in warm in order to minimize the cable length and capacitance.
The signal is then brought to CAEN V1729 ADC's and read out by a computer.
The sampling period of the ADC is set to 1.01~$\mu$s and the acquisition time window to 8274~$\mu$s.
128 channels are read out in total.
With an ADC resolution of 14~bits, the file size for one event amounts to about 2MB.

A photomultiplier (PMT) system is used to detect the prompt scintillation light emitted by charged particles transversing the TPC.
This defines the starting time for the charge drift and also provides a trigger for the data acquisition system.
The system consists of two PMTs (A1, A2) installed above the TPC in liquid argon.
The scintillation light of liquid argon lies in the UV-range and therefore the PMTs are coated with a layer of tetraphenyl-butadiene (TPB) serving as a wavelength shifter to blue light.
An external trigger system, consisting of one scintillator plate equipped with a PMT, is installed below the vessel (B1) to trigger on vertical muons; an additional one is installed on the side (B2) to trigger on oblique muons.
For cosmic muons the trigger logic was set to: A1\&A2\&B1 or A1\&A2\&B2.
More details on the wire planes, pre-amplification and DAQ can be found in earlier publications \cite{zeller,rossi}.

\subsection{UV laser}
\label{laser}
\begin{figure}
\centering	
\includegraphics[width=0.9\linewidth]{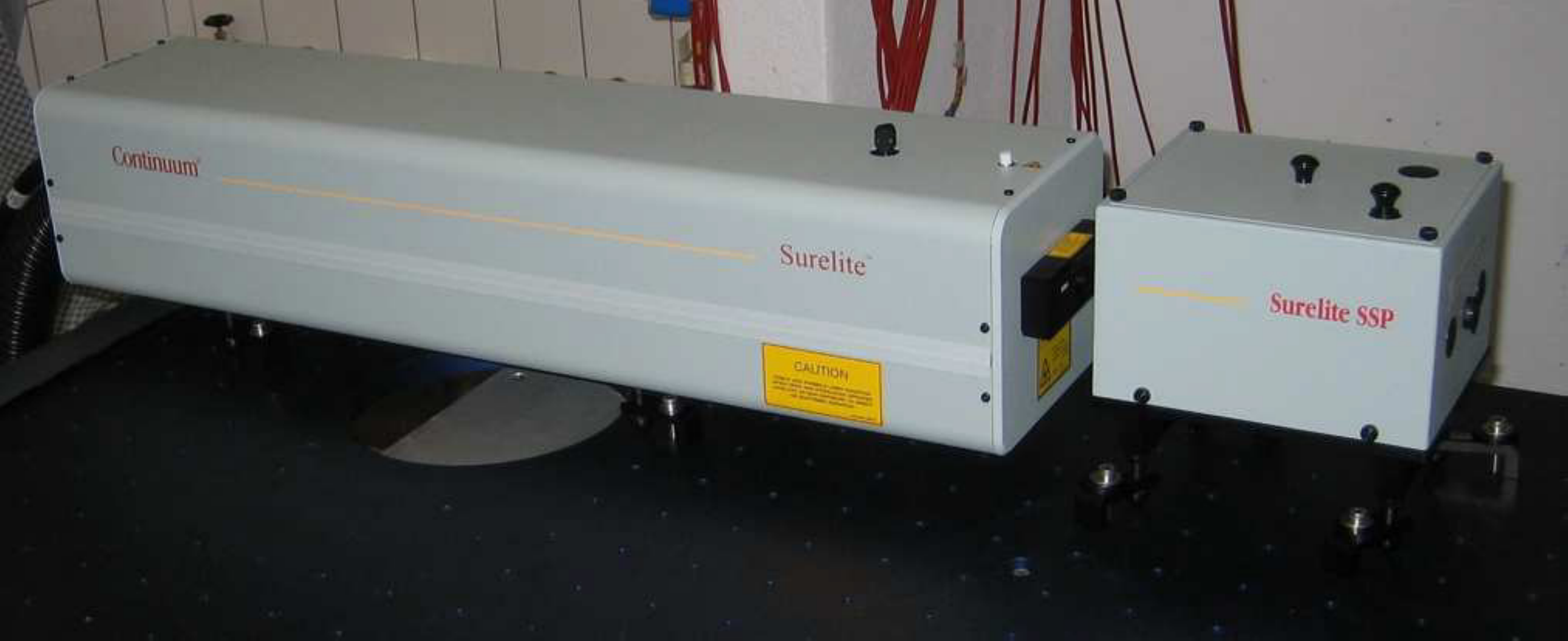}
\caption{Continuum Surelite Nd:YAG laser with a Surelite Separation Package (SSP) which separates the 266~nm laser beam from all the other harmonics.}
\label{fig:Laser}
\end{figure}

In order to measure the detector performance parameters, such as the liquid argon purity, the drift speed and the uniformity of the electric field, we use straight ionization tracks produced by UV laser beams.
Since the ionization potential in liquid argon is 13.84~eV, a laser with a wavelength of 89.8~nm would be needed for direct ionization.
Such a laser is not available on the market.
Moreover, the attenuation in air would complicate the experimental setup significantly.
However, a UV laser with a much longer wavelength of 266~nm but with very high intensity can be used to ionize the argon via a multi-photon absorption process.
Details of this approach are described in \cite{rossi, badhrees}.
Such lasers are readily available and the beam pulses can be transferred from the laser generator into the detector through air.
In Figure \ref{fig:Laser} the Nd:YAG laser\footnote[4]{Model: Surelite I-10,Continuum, 3150 Central Expressway, Santa Clara, CA 95051, USA.} used for ARGONTUBE is shown.
It generates a beam with a wavelength of 1064~nm and its harmonics at 555~nm, 332~nm and 266~nm.
The latter is selected with a wavelength separation unit (Surelite Separation Package (SSP)).
At this wavelength the pulses have an energy of 4~mJ and a duration from 4 to 6~ns; the beam divergence is less than 0.5~mrad.
The laser setup is mounted on a table placed above the detector.
In order to avoid non-uniform reflection/refraction at the surface of the liquid and to preserve beam geometry the laser beam is routed into the liquid argon via a quartz-glass feedthrough\cite{rossi} mounted on the cryostat top flange. 
The laser path goes from the top to the bottom of the TPC with an inclination of $3^{\circ}$ with respect to the vertical axis.
To trigger on laser signals a fast photo sensor is installed next to the laser head.
This sensor is sensitive enough to produce a signal from the parasitic reflections at the beam line optics.

\section{Operation and results}
\label{results}
\subsection{Operation of the detector}

Before filling the detector with liquid argon, the inner vessel is evacuated for about one week to minimize the amount of residual impurities.
Typically, a residual pressure of about $5 \times 10^{-5}$~mbar is reached.
Both vessels are filled up to 20~cm below the top flange, with the filling procedure taking about 9 hours.
Altogether 2700~liter of liquid argon are used for both vessels.
The argon filled to the inner vessel is purified by an oxygen trap filter.
During the whole operation of the detector the argon level, the pressure in the gas phase and the evaporation rate are constantly monitored.
The temperature in the liquid is measured as well with PT100 sensors at each meter from top to bottom.
After the whole system is cooled down to liquid argon temperature, the evaporation rate in the inner vessel is about 1 to 3~liter of gas per minute.
In the outer vessel the evaporation rate is around 100~liters of gas per minute.
The outer volume is periodically refilled to keep the liquid level above the one in the inner vessel.
The  inner volume is kept at an overpressure of 0.1~bar with respect to atmospheric pressure while the outer volume is open to atmospheric pressure.
At the top of the inner vessel the liquid argon is at boiling temperature, while at the bottom the temperature is by 0.3~$^\circ$C higher.
Because of the hydrostatic pressure of about 0.7~bar at 5~meters below the surface the liquid argon temperature at the bottom is still well below the boiling point.  

After filling, the liquid argon is recirculated through the oxygen-trapping filters with a recirculation speed of about 120~liters of liquid per hour.
The purification lasts for about 24 hours to let the argon reach the design level of impurities of about 1~ppm before the high voltage is ramped up.
During the measurement period the recirculation system is continuously running to get the best possible purity.

To apply the drift field, the input voltage of the Greinacher circuit is slowly ramped up to 0.96~kV, which corresponds to an expected cathode voltage of 120~kV and a drift field of 240~V/cm.
In these conditions the operation of the detector has shown to be very stable.
The Greinacher circuit input AC voltage is turned off during data taking to minimize interferences on the readout system.
The self-discharge of the Greinacher circuit is found to be less than 1\% per hour, therefore every 2-3 hours the circuit is recharged by turning on the AC input for several minutes.
Attempts to raise the drift field above 240~V/cm resulted in electric breakdowns at about 1.2~kV of input AC voltage, which corresponds to 150~kV an the cathode.
The exact cause of HV breakdowns is still under investigation.

\subsection{First tracks from a 5~m drift TPC}

\begin{figure}
\centering
\includegraphics[width=1\linewidth]{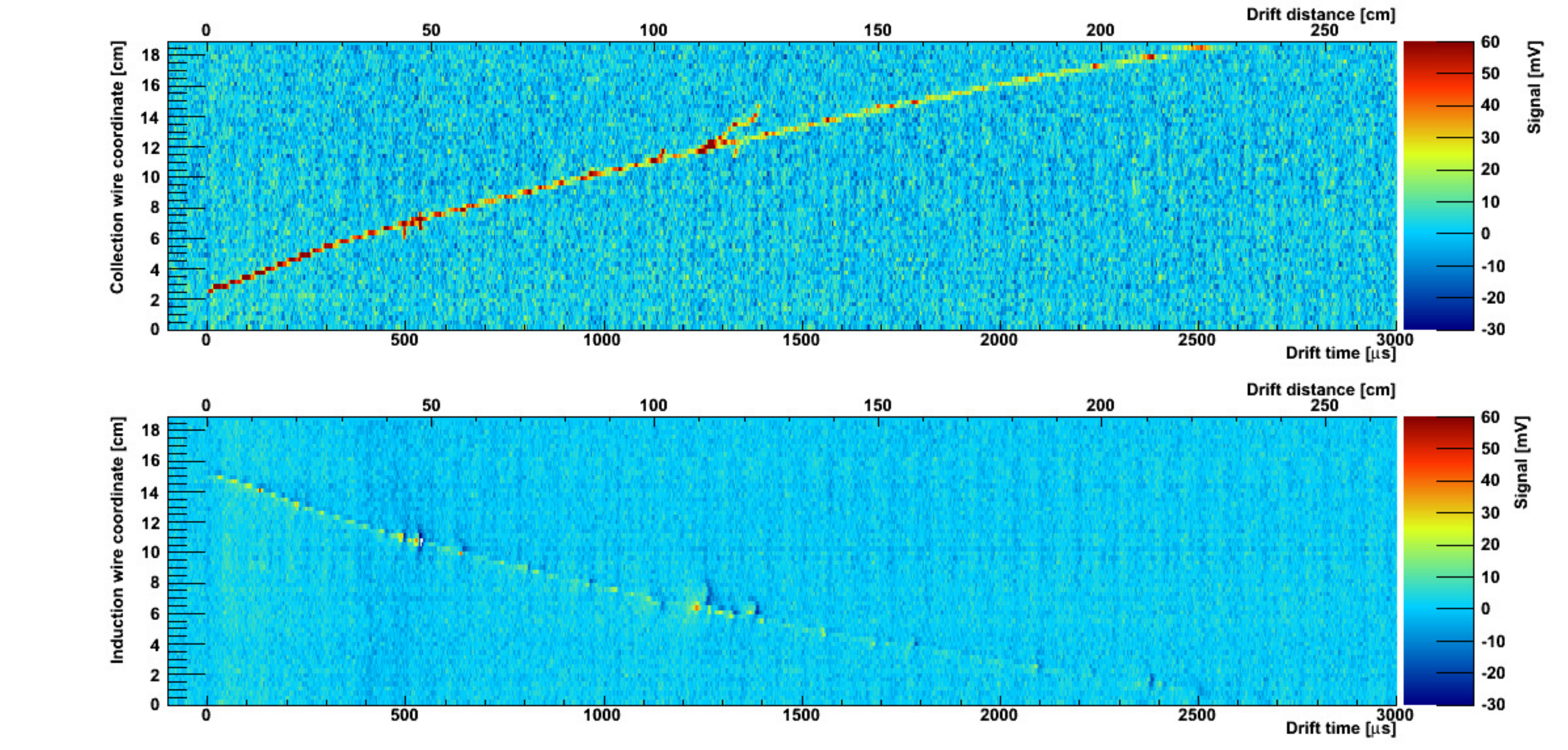} 
\caption{Cosmic muon entering the chamber with 2.2 m of drift distance. The upper picture shows the collection view and the lower picture shows the induction view. For a more detailed description of the figure see the text.}
\label{muon2}
\end{figure}

\begin{figure}
\centering
\includegraphics[width=1\linewidth]{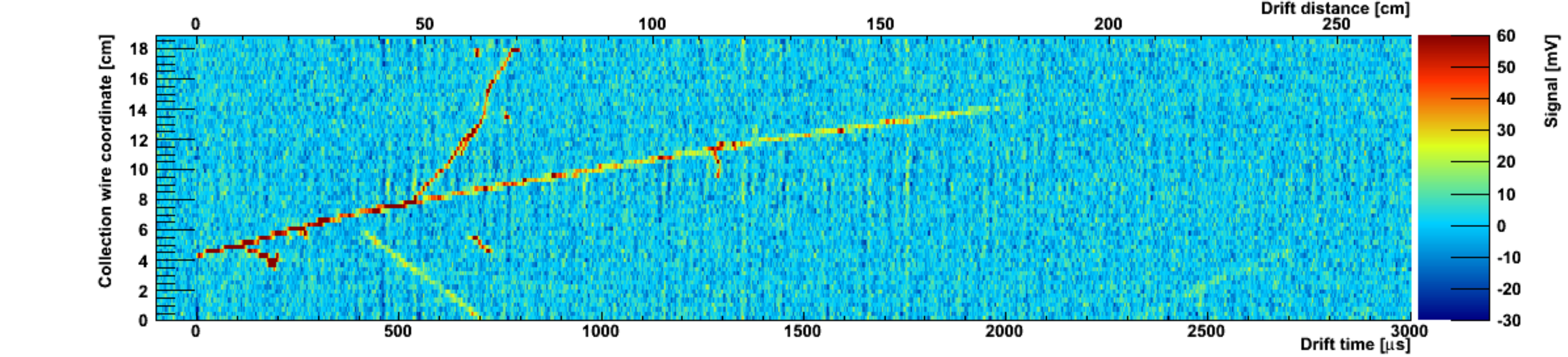} 
\caption{Collection view of a cosmic ray muon inducing delta-rays (see the text for details).}
\label{muon3}
\end{figure}

\begin{figure}
\centering
\includegraphics[width=1\linewidth]{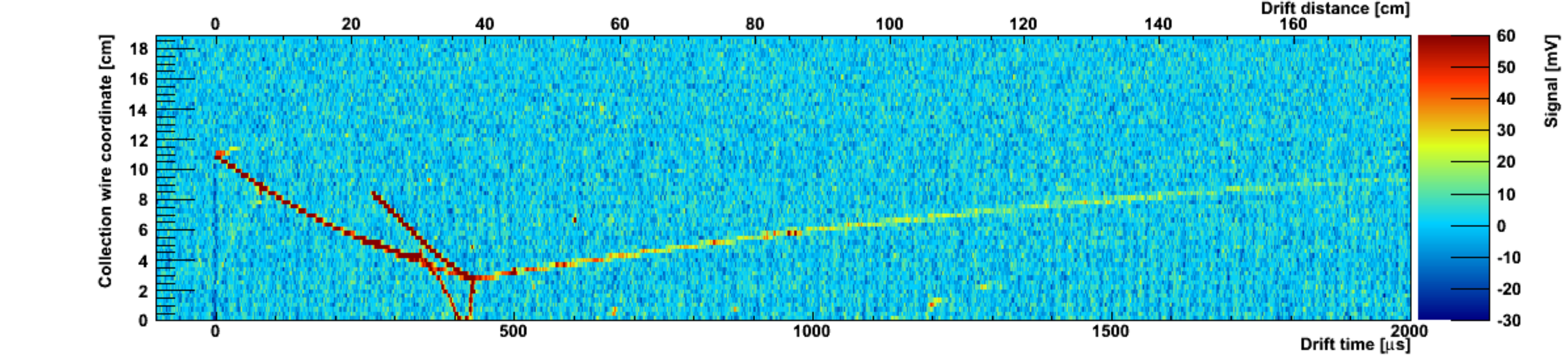} 
\caption{Collection view of a cosmic particle interacting with a nucleus with 3 secondary particles (see the text for details).}
\label{muon4}
\end{figure}

\begin{figure}
\centering
\includegraphics[width=1\linewidth]{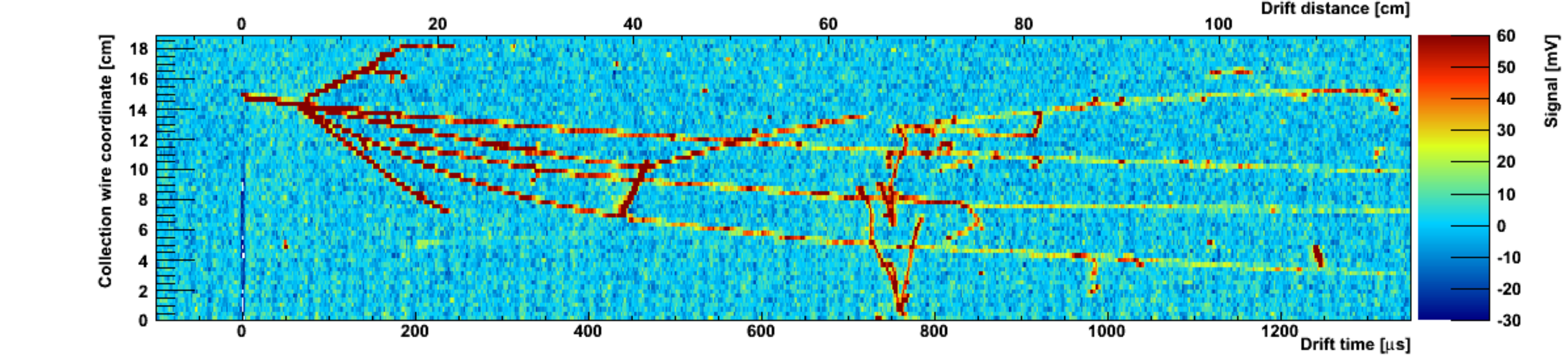} 
\caption{Collection view of a cosmic particle interacting with a nucleus with at least 6 secondary particles (see the text for details).}
\label{muon5}
\end{figure}

\begin{figure}
\centering
\includegraphics[width=1\linewidth]{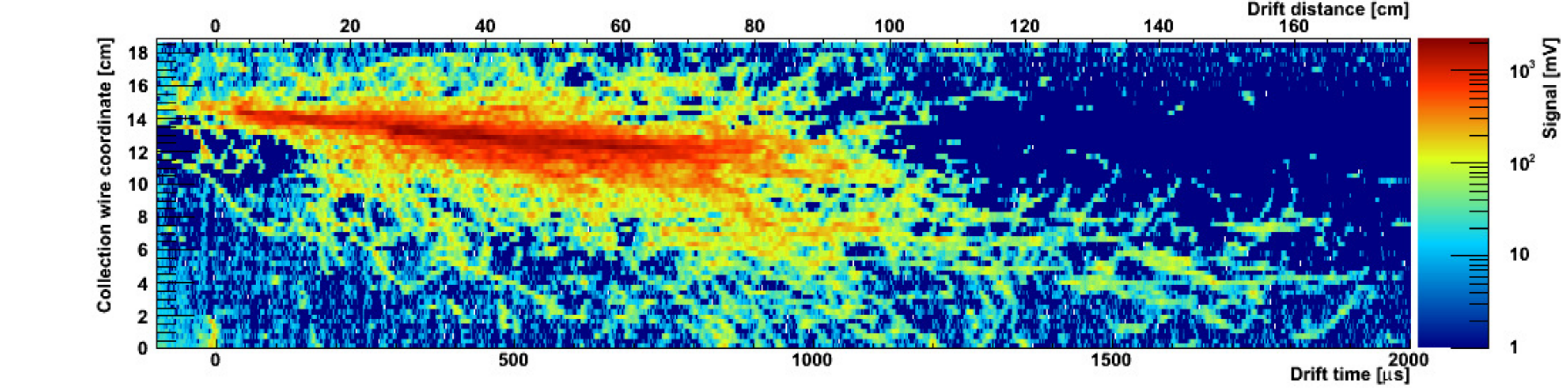} 
\caption{Collection view of an electromagnetic shower. The charge deposited is represented in logarithmic color scale (see the text for details).}
\label{shower1}
\end{figure}

\begin{figure}
\centering
\includegraphics[width=1\linewidth]{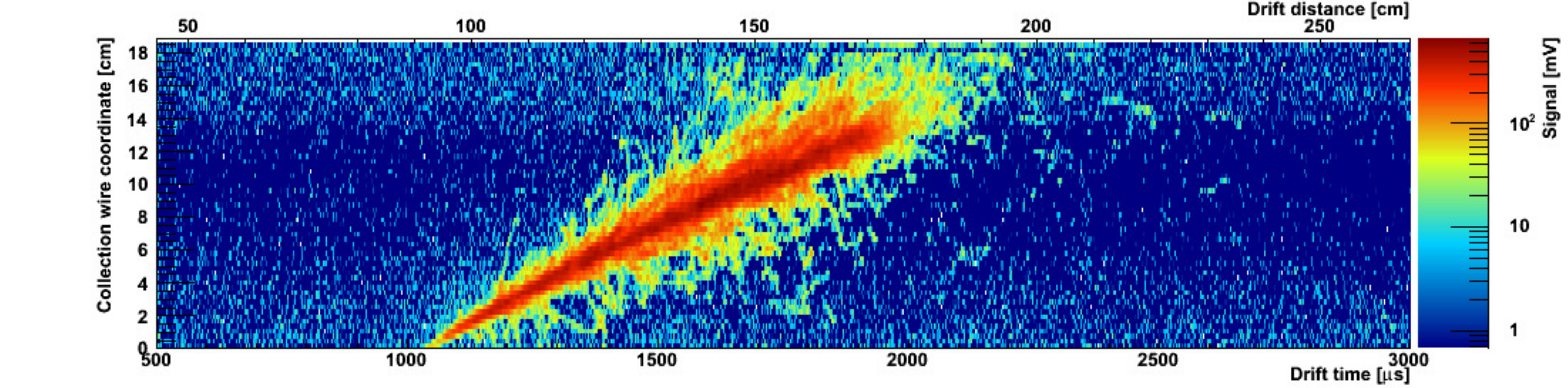} 
\caption{Collection view of an electromagnetic shower. The charge deposited is represented in logarithmic color scale (see the text for details).}
\label{shower2}
\end{figure}

\begin{figure}
\centering
\includegraphics[width=1\linewidth]{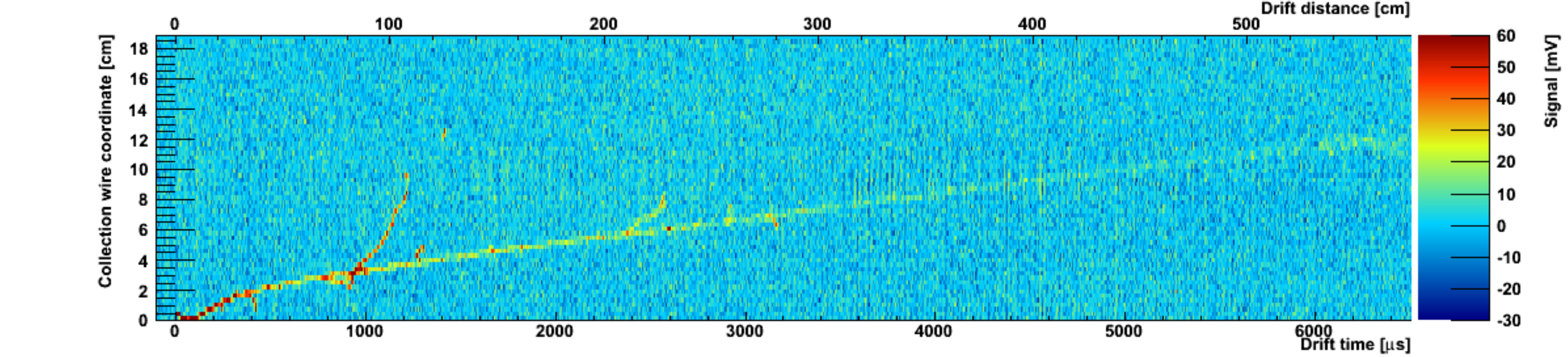} 
\caption{Collection view of a 5 m long muon track with up to 6.2~ms drift distance (see the text for details).}
\label{long_muon1}
\end{figure}

\begin{figure}
\centering
\includegraphics[width=1\linewidth]{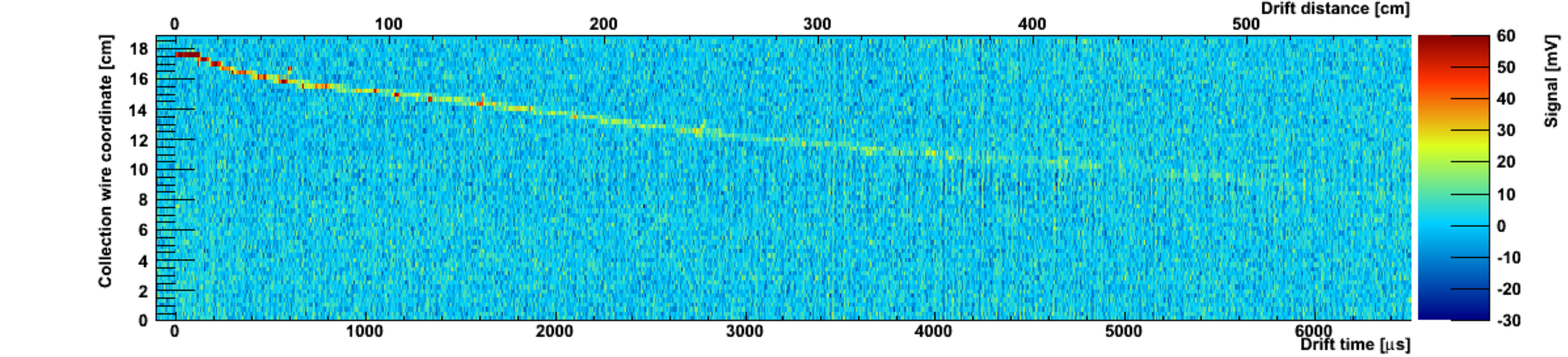} 
\caption{Collection view of a 5 m long muon track with up to 6.2~ms drift distance (see the text for details).}
\label{long_muon2}
\end{figure}

\begin{figure}
\centering
\includegraphics[width=1\linewidth]{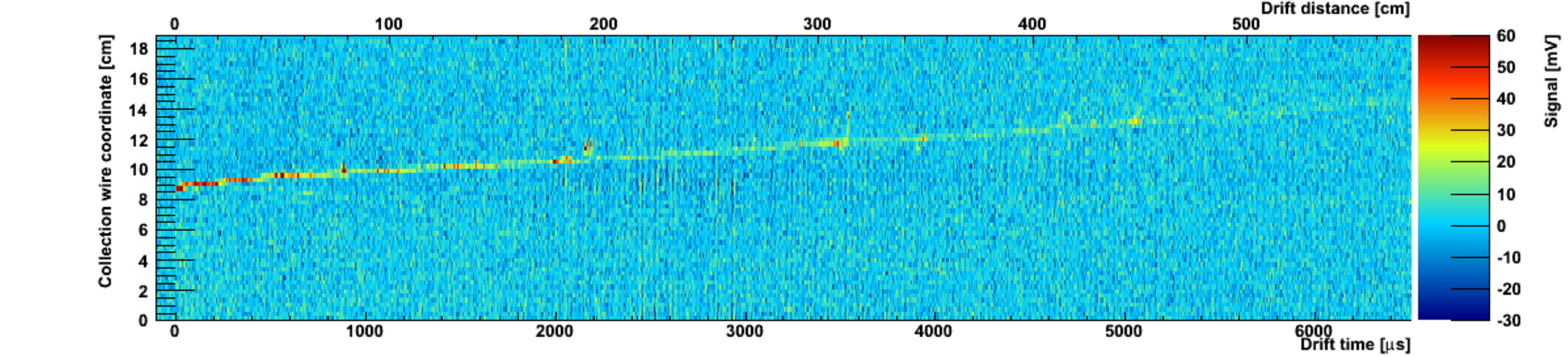} 
\caption{Collection view of a 5~m long muon track with up to 6.2 ms drift distance (see the text for details).}
\label{long_muon3}
\end{figure}

In a run performed in 2012 7'000 events from cosmic ray muons and 3'000 laser beam tracks were collected.
In Figures \ref{muon2} to \ref{long_muon3} a collection of cosmic ray induced tracks is shown.
Figure \ref{muon2} depicts a muon track.
The vertical axis corresponds to the readout plane coordinate in centimeters (3~mm wire pitch) and the horizontal axis corresponds to the drift time on the lower axis and to the drift distance on the upper axis.
The Figures are therefore turned by 90 degrees compared to the detector orientation in the laboratory.
Also the aspect ratio is compressed in time (distance) to fit the page width.
The charge deposited by the particle is represented with a color scale.
For these Figures the drift field in the TPC was set to 240~V/cm.
The drift distance is calibrated with the drift speed of 0.886~mm/$\mu$s calculated from the drift field of 240~V/cm \cite{drift}.
Figure \ref{muon3} shows a muon that induces several delta-rays, the longest delta-ray track being about 25 cm long.
In Figures  \ref{muon4} and \ref{muon5} a cosmic particle interacts with a nucleus, creating secondary particles.

The LArTPC technology shows as well good capabilities  to measure electromagnetic showers.
In Figures \ref{shower1} and \ref{shower2} almost fully contained electromagnetic showers are shown.
The charge deposited is represented with a logarithmic color scale due to the high charge deposition of an electromagnetic shower. 
In Figures \ref{long_muon1}, \ref{long_muon2} and \ref{long_muon3} muons passing whole the detector from the top to the bottom are shown. They show tracks with about 5 meters drift distance, which corresponds to a maximum charge drift time of 6.2~ms.
The end of the tracks are visible but rather weak due to the purity and to the low drift field at the time those events were recorded. At small drift distance the signal from cosmic muons entering nearly normal to the readout wires give a signal of about 40-50~mV, resulting in a signal-to-noise ratio (S/N) of about 10.

\begin{figure}
\centering
\includegraphics[width=1\linewidth]{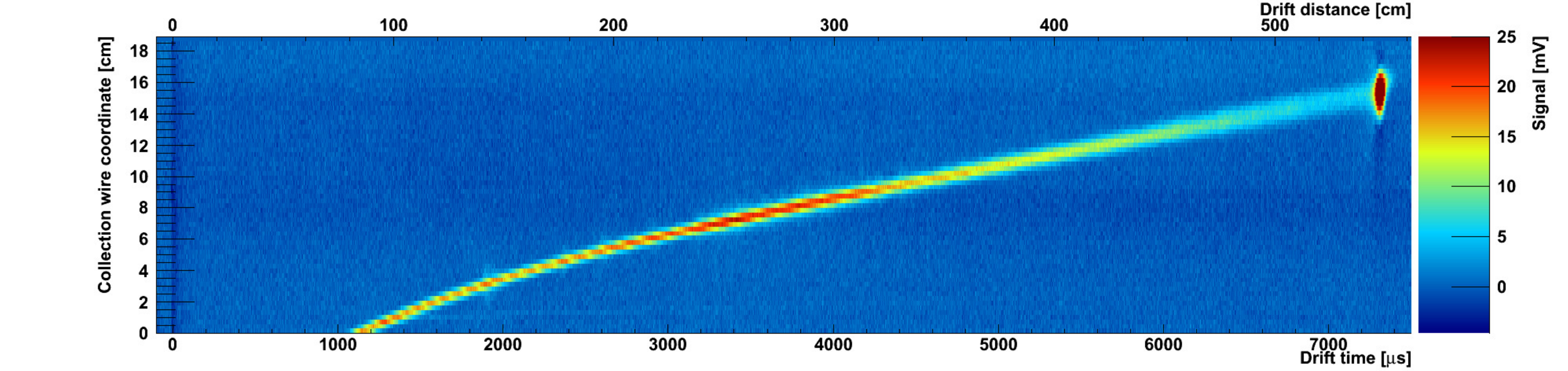} 
\caption{100 superimposed laser tracks with 5~m drift distance for a drifting time of 7.4~ms. A voltage of 100~kV provides the drift field of 200~V/cm (see the text for details).}
\label{laser1}
\end{figure}

\begin{figure}
\centering
\includegraphics[width=1\linewidth]{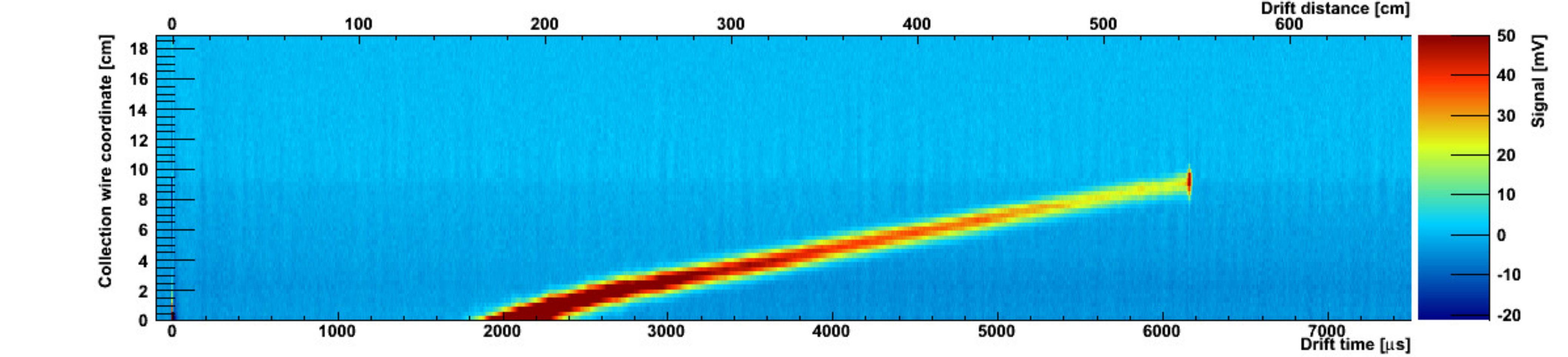} 
\caption{100 superimposed laser tracks with 5~m drift distance for a the drifting time is 6.2~ms. A voltage of 120~kV provides the drift field of 240~V/cm (see the text for details).}
\label{laser2}
\end{figure}

In Figures \ref{laser1} and \ref{laser2} UV-laser induced signals are shown.
The laser beam enters from the side and goes to the cathode, where a cloud of electrons is produced due to photoelectric effect on the gold-coated cathode surface.
100 laser tracks are superimposed to get much better signal to noise ratio in these Figures.
Therefore, the 5~m  drift is much better visible compared to single cosmic ray events.
Assuming a homogeneous drift field we measure the end of the track from the photoelectric emission at the cathode to be at 5.4m instead of 5.0~m.
Since we do not have enough information for a correction, we thus assign a systematic uncertainty to the drift field of 8\%.

\subsection{Charge life time and argon purity}

The charge $q$ collected at the TPC wires after a drift distance $x$ is defined by:

\begin{equation}
q=q_0e^{\frac{-t}{\tau}} \ , \ \ t=\frac{x}{v}
\end{equation}

where $q_0$ is the charge left after the primary recombination, $t$ is the drift time, $\tau$ is the charge lifetime, which is a function of the purity and temperature, and $v$ is the drift velocity. 
The drift time is defined by the trigger time given by the PMTs and the arrival time of the electrons at the wire planes.

\begin{figure}
\centering
\includegraphics[width=0.7\linewidth]{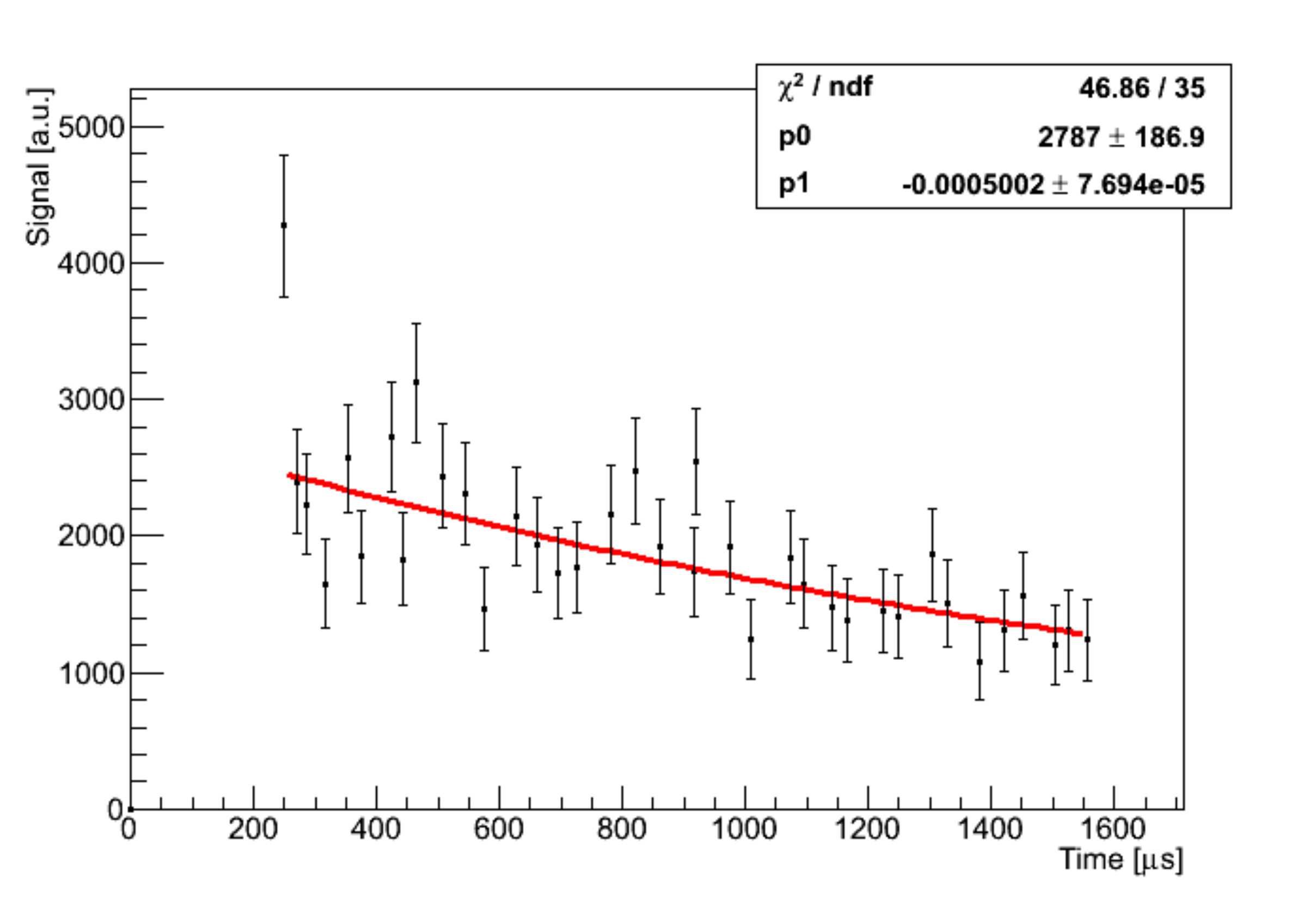} 
\caption{Charge attenuation of a recorded muon track versus the drift time. The charge lifetime is found to be $\tau$~=~2.00$\pm$0.31~ms. The fitting function is shown in equation~(2.1) where q$_{0}$ is  p0 and -1/$\tau$ is p1.}
\label{Purityfit}
\end{figure}

\begin{figure}
\centering
\includegraphics[width=0.7\linewidth]{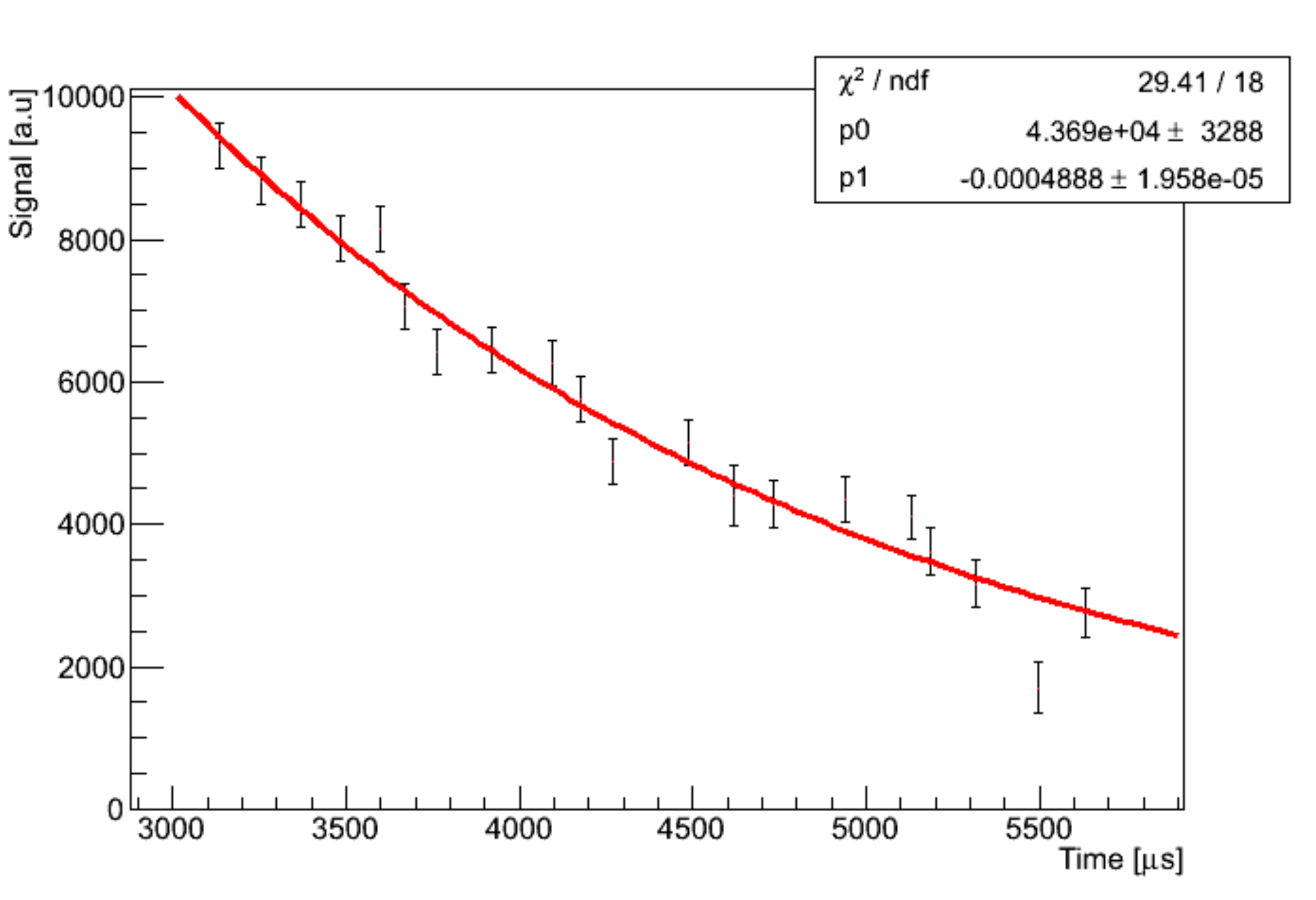} 
\caption{Charge attenuation of 100 laser tracks versus the drift time. The charge lifetime is found to be $\tau$~=~2.05$\pm$0.08~ms. The fitting function is shown in equation~(2.1) where q$_{0}$ is  p0 and -1/$\tau$ is p1.}
\label{Purityfit2}
\end{figure}

The liquid argon purity in the TPC was measured using long inclined cosmic muon and laser beam tracks by analyzing the exponential decay of the charge signal versus the drift time.
For the reconstruction a tracking algorithm based on Hough-transformation was implemented.
In Figure~\ref{Purityfit} the collected charge as a function of drift time along a muon track is shown.
An event similar to the one shown in Figure~\ref{muon2} was used. 
A fit to this particular event results in a charge lifetime of (2.0$\pm$0.3)~ms.
The collected charge from the 100 laser tracks shown in Figure~\ref{laser1} is given as a function of the drift time in Figure~\ref{Purityfit2}.
A charge lifetime of (2.11$\pm$0.05)~ms is found.
The results of the two different methods are compatible.
Note, however, the much smaller uncertainty due to the statistical gain for the laser measurement.
These measurements where performed after purifying the liquid argon for about 48 hours, which results in about 0.15~ppb oxygen equivalent impurities concentration.
Immediately after the filling through the cryogenic oxygen trapping filters we measure a charge lifetime of only about 0.5~ms.

The detailed analysis of the ARGONTUBE performance including studies of charge diffusion, coordinate and energy resolution and the determination of the ultimate reachable argon purity will be the subject of an upcoming publication.

\section{Conclusions}

We reported on experimental results from ARGONTUBE, a liquid argon TPC with up to 5 m long drift distance.
We successfully operated the detector with its vacuum and cryogenic systems, the liquid purification system, the Greinacher high voltage system and the laser calibration system.
We presented the first muon tracks and laser beam induced tracks ever measured with a 5~m long drift length.
A maximum cathode-anode voltage of 150 kV was reached, while a stable operating voltage of 125 kV (corresponding to a drift field of 250~V/cm) was established over many days of continuous operation.
The liquid argon purity we achieved was better than 0.15~ppb, which corresponds to a charge lifetime of 2~ms, determined from cosmic muons and laser track measurements.


\begin{thebibliography}{99}
\expandafter\ifx\csname url\endcsname\relax
  \def\url#1{\texttt{#1}}\fi
\expandafter\ifx\csname urlprefix\endcsname\relax\def\urlprefix{URL }\fi

\bibitem{PDG}
J. Beringer et al. (Particle Data Group), \textit{The Review of Particle Physics, Phys. Rev.} \textbf{D 86} (2012) 010001.

\bibitem{pontecorvo} 
B. Pontecorvo, \textit{ Mesonium and anti-mesonium, Sov.Phys.} \textbf{JETP6} (1957)429  and  \textit{Sov.Phys.} \textbf{JETP7}  (1958) 172.

\bibitem{MNS}
Z. Maki, M. Nakagawa, and S. Sakata, \textit{Remarks on the Unified Model of Elementary Particles, Prog. Theor. Phys.} \textbf{28} (1962) 870.

\bibitem{Rubbia}
C. Rubbia, \textit{The liquid argon time projection chamber: a new concept for neutrino
detectors}, CERN-EP-INT-77-8, 16 May 1997.

\bibitem{ICARUSproposal}
F. Arneodo et al., [ICARUS Collab.],  \textit{The ICARUS Experiment, A Second-Generation Proton Decay Experiment and Neutrino Observatory at the Gran Sasso Laboratory}, arXiv:0103008 (hep-ex) (2001).

\bibitem{GLACIER1}
A. Rubbia, \textit{Experiments For CP-Violation: A Giant Liquid Argon Scintillation, Cerenkov And Charge Imaging Experiment ?}, arXiv:hep-ph/0402110.

\bibitem{GLACIER2}
A. Ereditato and A. Rubbia, \textit{Ideas for future liquid Argon detectors,
Nucl. Phys. Proc. Suppl.} \textbf{139} (2005) 301.

\bibitem{GLACIER3}
A. Ereditato and A. Rubbia, \textit{The liquid Argon TPC: a powerful detector for future neutrino experiments and proton decay searches,
Nucl. Phys. Proc. Suppl.} \textbf{154} (2006) 163.

\bibitem {zeller} 
M. Zeller et al., \textit{Ionization signals from electrons and alpha-particles in mixtures of liquid Argon and Nitrogen - perspectives on protons for Gamma Resonant Nuclear Absorption applications,
JINST} \textbf{5} (2010)  10009.

\bibitem {rossi} 
B. Rossi et al., \textit{A prototype liquid Argon Time Projection Chamber for the study of UV laser multi-photonic ionization,
JINST} \textbf{4} (2009)  07011.

\bibitem{larn}
A. Ereditato et al., \textit{Study of ionization signals in a TPC filled with a mixture of liquid Argon and Nitroge,
JINST} \textbf{3} (2008) 10002.

\bibitem {badhrees} 
I. Badhrees et al., \textit{Measurement of the two-photon absorption cross-section of liquid argon with a time projection chamber,
New J. Phys.} \textbf{12} (2010) 113024.

\bibitem{ICARUS} 
S. Amerio et al., [ICARUS Collab.], \textit{Design, construction and tests of the ICARUS T600 detector,
Nucl. Instrum. and Methods} \textbf{A 527} (2004) 329.

\bibitem{Imel}
J. Thomas and D. A. Imel, \textit{Recombination of electron-ion pairs in liquid argon and liquid xenon,
Phys. Rev.} \textbf{A 36} (1987) 614.

\bibitem{diffusion}
E. Shibamura, et al., \textit{Ratio of diffusion coefficient to mobility for electrons in liquid argon,
Phys. Rev.} \textbf{A 20} (1979) 2547.

\bibitem{diffusion2}
S.E. Derenzo, \textit{Electron diffusion and positive ion charge retention in liquid-filled high-resolution multistrip ionization-mode chambers},
Lawrence Berkeley Laboratory, Group A Physics Note No. 786 (1974).

\bibitem{diffusion3}
S.E. Derenzo et al., \textit{Test of a liquid argon ion chamber with a 20mm RMS resolution,
Nucl. Instrum. and Methods} \textbf{122} (1974) 319.

\bibitem{atrazhev}
V.M. Atrazhev and I.V. Timoshkin, \textit{Transport of Electrons in Atomic Liquids in High Electric Fields,
IEEE Trans. on Dielectrics and Electrical Insulation} \textbf{5} (1998) 450.

\bibitem{swan}
D.W. Swan and T.J. Lewis, \textit{The Influence of Cathode and Anode Surfaces on the Electric Strength of Liquid Argon,
Proc. Phys. Soc.} \textbf{78} (1961) 448.

\bibitem{drift}
W. Walkowiak, \textit{Drift velocity of free electrons in liquid argon,
Nucl. Instrum. and Methods} \textbf{A 449} (2000) 288.

\end{thebibliography}
\end{document}